\documentclass[aps,prl,twocolumn,groupedaddress,nofootinbib,superscriptaddress]{revtex4}

\usepackage{amsmath,amssymb,epsfig,ifthen,capt-of,calc}

\newcommand{\bignone}{}
\newcommand{\nonesep}{}
\newcommand{\tmem}[1]{{\em #1\/}}
\newcommand{\tmop}[1]{\ensuremath{\operatorname{#1}}}
\newcommand{\tmsamp}[1]{\textsf{#1}}
\newcommand{\tmtextrm}[1]{{\rmfamily{#1}}}

\begin{document}

\title{Benchmarks for new strong interactions at the LHC}\author{J. Hirn$^1$, A.
Martin}\affiliation{Department of Physics, Yale University, New Haven CT 06520}  \author{V. Sanz} \affiliation{Department of Physics, Boston University, Boston MA 02215}

\begin{abstract}
  New strong interactions at the LHC may exhibit a richer structure than
  expected from simply rescaling QCD to the electroweak scale. In fact, a
  departure from rescaled{\tmsamp{}} QCD is required for compatibility with
  electroweak constraints. To navigate the space of possible scenarios, we use
  a simple framework, based on a 5D model with modifications of AdS geometry
  in the infrared. In the parameter space, we select two points with
  particularly interesting phenomenology. For these benchmark points, we
  explore the discovery of triplets of vector and axial resonances at the LHC. 
\end{abstract}

\maketitle

  {\bf Introduction:}- The two main experimental collaborations at LHC
  -ATLAS and CMS- classify models of electroweak symmetry breaking (EWSB) in
  two groups: Supersymmetry (SUSY) or Exotics. The two groups are on a
  different footing, though: detailed studies of SUSY phenomenology abound,
  whereas the collider phenomenology of most models referred to as Exotics is
  still in its infancy, due mostly to the intricacies in handling strong
  coupling{\footnote{In practice, only one very specific model of strong
  interactions has been implemented in Monte-Carlo generators and studied at
  Tevatron: the straw-man model {\cite{Lane:1999uk}}.}}.
  
  The profusion of phenomenological studies of SUSY was spurred by the
  successes of the Minimal Supersymmetric Standard Model (MSSM)
  {\cite{Dimopoulos:1981zb}}, and the parametrical simplicity of minimal
  Supergravity (mSUGRA). mSUGRA-MSSM provided a compact, manageable framework
  appropriate for collider studies.
  
  We still lack a similar simplifying assumption to parameterize strong
  interactions. In this paper, we take a step to remedy this by constructing a
  flexible yet manageable description of resonance interactions, making use of
  the idea that extra-dimensional (ED) models provide a description of
  strongly-interacting theories.
  
  Exact results of such an equivalence between theories in different
  dimensions (the AdS/CFT correspondence
  {\cite{hep-th/9802109,hep-th/9802150}}) have only been obtained in
  particular cases. But for the purpose of LHC simulations, we do not need an
  exact duality to hold: we can study ED models whose essential properties are
  the same as those of strong interactions. This gives us a qualitative, and
  sometimes quantitative insight into strong interactions
  {\cite{hep-ph/0501128,hep-ph/0501218,Hirn:2005nr}}.
  
  Since electroweak measurements {\cite{Holdom:1990tc,Yao:2006px}} exclude a rescaled copy
  of QCD \ (Technicolor {\cite{Weinberg:1979bn,Susskind:1978ms}}), a departure
  from this rescaling is necessary. In the ED framework the departure can be
  described in terms of a few parameters. This is the idea behind Holographic
  Technicolor (HTC) {\cite{hep-ph/0606086}}.
  
  {\bf Charting the unmapped territory : } While HTC shares many features
  with purely-AdS models, such as approximate custodial $\tmop{SU} (2)$
  symmetry ($T \approx 0$), the freedom granted by non-AdS geometry allows for
  a suppression of the $S$ parameter. In a pure AdS model, $S$ can only be
  tamed by suppressing couplings between the resonance sector
  and the SM fermions (fermiophobic scenario) {\cite{hep-ph/0409126}}. In HTC,
  because of cancellations between the nearly degenerate states, $S$ can be
  small while maintaining a nonzero coupling of fermions to light resonances {\cite{Hirn:2006wg}}.
  This coupling allows $s$-channel production of resonances, observable in the
  early stages of the LHC. Conversely, in the fermiophobic case, discovery of
  the strongly-coupled sector is delayed to late stages of the LHC.
  
  In this paper, we confine ourselves to Dynamical EWSB without a Higgs, but
  with light ($\lesssim$1 TeV) spin-1 resonances coupled to the
  $W$'s{\footnote{As opposed to the DBESS model, whose resonances are
  approximately decoupled from the electroweak sector
  {\cite{Casalbuoni:1995qt}}.}}. Such light resonances can help unitarizing $W W$
  scattering while interacting weakly. We will consider only the lightest two
  triplets of resonances $(W_{1,
  2}^{\pm}, Z_{1, 2})$. 
An effective Lagrangian describing those resonances and their couplings to the SM would introduce
  ${\cal O} (100)$ new parameters. Using an ED description
  we drastically reduce  the number of parameters involved to just four. We do not model the fermions in ED but choose the fermion-resonance
  coupling $g_{f fV}$ as a free parameter{\footnote{Dealing with the
  intricacies of fermions in ED would provide an interesting picture of
  extended technicolor models {\cite{Rius:2001dd}}. }}.

  The triplets of resonances are described by
  ED gauge fields $\tmop{SU} (2)_L \otimes \tmop{SU} (2)_R$ propagating in a
  compact geometry given by $\tmop{ds}^2 = w (z)^2 (dx^2 - dz^2)$, where $l_0
  \leqslant z \leqslant l_1$. We define two effective warp factors
  {\cite{Hirn:2005vk,hep-ph/0606086}} $w_X = (l_0 / z) \exp \left( \frac{o_X}{2}  \left(
  \frac{z - l_0}{l_1} \right)^4 \right)$, $X = A, V$. The power in $(z /
  l_1)^4$ was based on walking technicolor arguments
  {\cite{Appelquist:1986an}}, but irrelevant for the LHC phenomenology:
  one can absorb the effect of a different power in the $o_X$ value. One can
  then extend minimally the setup to introduce $U (1)_{B - L}$ and choose
  boundary conditions that preserve just $U (1)_{\tmop{em}}$, leading to a
  massless photon, and very light $W, Z$. Pure AdS geometry corresponds to
  $o_X = 0$.
  
  We assume strong interactions are parity symmetric. Once coupled to
  the EW sector, physical mass eigenstates $(W_{1, 2}^{\pm}, Z_{1, 2})$ are an
  admixture of axial and vector eigenstates. Therefore, both triplets of resonances couple
  to the longitudinal $W, Z$.

  {\bf Benchmark points:} Our description is very economical in terms of
  new parameters: the size of the ED ($l_1$), the amount of departure from
  rescaled-QCD ($o_{V, A}$) and he coupling to fermions ($g_{f f V}$).
  
  We emphasize that HTC is not a model of EWSB, rather it is an
  organizational scheme which allows us to describe viable resonance models in
  terms of a few parameters. To give a sample of the phenomenology coming out of this description, we chose two benchmark points
  in the parameter space, HTC1 and HTC2.

  \begin{center}
    \label{table5}\begin{tabular}{|c|c|c|c|c|}
      \hline
      Signal  &  $\ell_1 (\tmop{TeV}^{- 1})$  &  $o_V$  &  $o_A$  & $g_{ffV}
      / g_2$ \\
      \hline
      HTC 1  &  6.3 &  -10  &  0  & $0.1$\\
      \hline
      HTC2 & 8 & -22.5 & 0 & 0.05 \\
      \hline
    \end{tabular}
  \end{center}

  For the point HTC1 above, $M_{W_{1, 2}} \sim (600, 680) \tmop{GeV}$ and
  width $\Gamma_{W_{1, 2}} \sim (4, 2) \tmop{GeV}$. For HTC2, $M_{W_{1, 2}}
  \sim (500, 630) \tmop{GeV}$ and width $\Gamma_{W_{1, 2}} \sim (1, 4)
  \tmop{GeV}$.   Small $\frac{\Gamma}{M} $ can be understood from a purely 4D point of view: HTC1 and HTC2 describe
  resonances as bound states of a strongly coupled theory, but whose
  interactions are determined by the number of colors, $N_{\tmop{TC}},$ of the
  strong sector. Large values of $N_{\tmop{TC}}$ correspond to weakly coupled
  (i.e. narrow) resonances. In the HTC1 point, both resonances are visible in the $s$-channel
  production to $W Z$ and $W \gamma$. The mass separation between the
  $W_1$ and $W_2$ is larger in HTC2, leading to a different
  phenomenology: only the lightest resonance is visible in the $s$-channel
  production to $W \gamma$.
  
  {\bf Constraints : } The geometry parameters $o_V, o_A$ $l_1$ are
  constrained by LEP limits on anomalous triboson couplings
  {\cite{Yao:2006px}}. The $g_{f fV}$ are constrained by direct $Z', W'$
  Tevatron cross section bounds {\cite{:2007sb,hep-ex/0611022}} and
  by contact interaction limits {\cite{hep-ph/0106251,Yao:2006px}}. We have
  also checked that the resonances do not disrupt the measured
  Tevatron $ W\, Z$, $\gamma W$ cross sections
  {\cite{hep-ex/0503048,hep-ex/0702027}} and high $p_T$ distibutions
  {\cite{hep-ex/0503048,CDFnote}}.
  
  {\bf $s$-channel production to $W Z$: } We first consider $\hat{s}$-channel
  production of a new vector resonance to $W^{\pm} Z$ final state. Within the
  narrow width approximation (NWA), the signal cross section for each new resonance
  is \
  \begin{eqnarray}
    \left(  \frac{1}{s}  \frac{d\mathcal{L}}{d \tau}  \right)  \frac{g^2_{f f
    V} g^2_1 M^5_{W_{1, 2}}}{1152 M^2_W M^2_Z \Gamma_{W_{1, 2}}}  \left( 1
    +\mathcal{O} \left( \frac{M^2_W}{M^2_{W_{1, 2}}} \right) \right) & , & 
  \end{eqnarray}
  \label{eq:NWArho}where $s$ is the square of the LHC center of mass energy, $g_1$ is a triboson coupling and
  $\frac{d\mathcal{L}}{d \tau} = \int \bignone \frac{dx_1}{x_1} f_q (x_1)
  f_{\bar{q}} ( \frac{M^2_{\rho}}{x_1 s})$ {\cite{Eichten:1984eu}}. 
  
  The fully leptonic decay mode $W^{\pm} Z \rightarrow 3 \ell + \nu$, $\ell=e,\mu$, is the
  cleanest mode and is not plagued by difficult QCD backgrounds. The important
  backgrounds for this process are, $W^{\pm} Z \rightarrow 3 \ell + \nu$
  (irreducible), $ZZ \rightarrow 4 \ell$, $Z + b \bar{b} \rightarrow \ell^+
  \ell^- + b \bar{b}$ and $t \bar{t}$. All of the backgrounds were generated
  at parton level using ALPGENv13~{\cite{Mangano:2002ea}}.

  We implemented HTC into the event generator MadGraph {\cite{Alwall:2007st}}
  and its add-on BRIDGE {\cite{hep-ph/0703031}}. We modified both programs 
  to handle anomalous triboson vertices.  The parton level events were passed
  through PYTHIAv6.4 {\cite{hep-ph/0603175}} for parton showering,
  fragmentation, and hadronization, and then through PGS 4.0~{\cite{PGS}} for
  detector simulation{\footnote{We use the PGS ATLAS parameter set available
  with MADGRAPH4.0. The relevant parameters are the calorimeter segmentation
  $\Delta \eta \times \Delta \phi = 0.1 \times 0.1$, the jet energy resolution
  $\delta E / E_{\tmop{jet}} = 0.8 / \sqrt{E}$, and the electromagnetic
  resolution $\delta E / E_{\tmop{em}} = 0.1 / \sqrt{E} + 0.01$.}}. 
  
  The standard minimal cuts we impose are: {\tmem{1.)}} exactly $3$ leptons
  with $p_T > 10 \tmop{GeV}, | \eta | < 2.5$. Of these leptons, at least one
  must have $p_T > 30 \tmop{GeV}$, {\tmem{2.) }}2 same-flavor, opposite charge
  leptons reconstruct the $Z$ mass to within $3 \Gamma_Z$, and {\tmem{3.)
  }}$H_{T, jets} < 125 \tmop{GeV}$, where $H_{T, jets} = \sum_{jets} p_{T,
  jets}$. Cut {\tmem{1.)}} reduces the background from $ZZ$, while \ cuts
  {\tmem{2.)}} and {\tmem{3.)}} suppress the contribution from $t \bar{t}$.
  The significance can be enhanced by cutting on the
  minimum $p_T$ of the $W$ and $Z$ ($p_T > 100 \tmop{GeV}$). By assuming
  $E_{T, \tmop{miss}} = p_{T, \nu} $ and constraining $(p_e + p_{\nu})^2 =
  M^2_W$ we can solve for the $\hat{z}$ momentum of the
neutrino{\footnote{There is a two-fold ambiguity in $p_{z, \nu}$
  which we resolve by taking the solution with greater $\widehat{p_{}}_{\ell}
  \cdot \widehat{p_{}} _{\nu}$}}. This allows us to reconstruct the $W$
  momentum. See TABLE I for details.
  
  \begin{table}[h]
    $\begin{array}{|c|c|c|c|}
      \hline
      \text{\tmtextrm{Process}} & \sigma_{0, LO} &
      \epsilon_{} (\%) & N_{EV} ( \mathcal{L} = 10 \tmop{fb}^{- 1})\\
      \hline
      \tmop{HTC} 1 (W Z) & 0.06 \tmop{pb} & 51 & 285\\
      \hline
      \tmop{HTC} 2 (W Z) & 0.02 \tmop{pb} & 48.5 & 97\\
      \hline
      W^{\pm} Z \rightarrow 3 \ell + \nu & 0.965 \tmop{pb} & 1.82 & 177\\
      \hline
      ZZ \rightarrow 4 \ell & 0.116 \tmop{pb} & 1.56 & 18\\
      \hline
      Zb \bar{b} & 11.4 \tmop{pb} & 5.17 \times 10^{- 3} & 6\\
      \hline
      t \bar{t} (\tmop{leptonic}) & 22.8 \tmop{pb} & 4.60 \times 10^{- 3} &
      10\\
      \hline
    \end{array}$ \label{table:wz}
    \caption{Processes, cross sections and efficiencies in the $W Z
    \rightarrow 3 \ell + \nu$ mode after a cut on $p_T > 100 \tmop{GeV}$.}
  \end{table}
  
  \begin{figure}[t]
    \epsfig{file=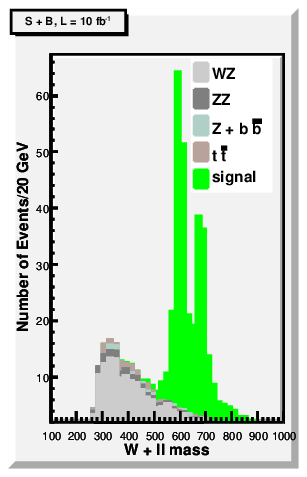,width=1.5in,height=1.25in}\epsfig{file=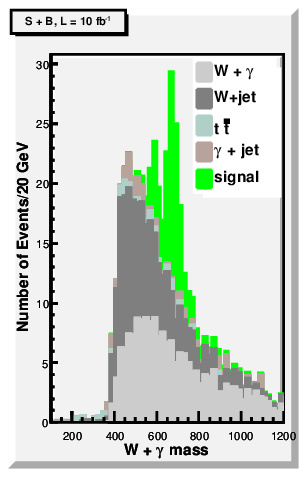,width=1.5in,height=1.25in}
    
    \epsfig{file=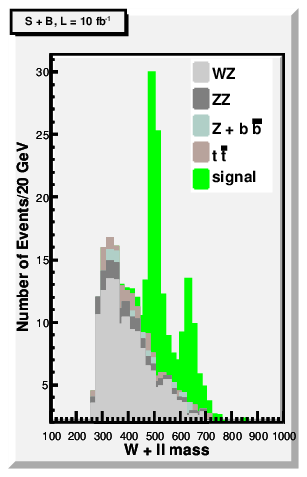,width=1.5in,height=1.25in}\epsfig{file=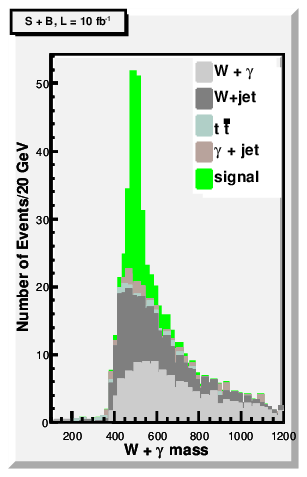,width=1.5in,height=1.25in}
    
    \label{DYfig}
    \caption{HTC1 (upper) and HTC2 (lower) - $W Z$ (left) and $W \gamma$
    (right) channels\, ($\mathcal{L}= 10 \tmop{fb}^{- 1}$). }
  \end{figure}
  Throughout this paper, cross sections include branching ratios to $\ell=e,\mu$ for signal and $t \bar{t}$ and $\ell=e,\mu, \tau$ for the other backgrounds.
  Detector effects, such as smearing and imperfect particle identification,
  are included in the efficiency quoted above. The most important detector
  effect in this channel is the lepton identification efficiency, $\sim 85\%$
  in the kinematic region of interest.
  
  Values of $g_{ffV}$ in HTC1,2 are compatible with TeVatron-LEP limits and
  still both peaks would be discovered within the first few $\tmop{fb}^{- 1}$
  at the LHC{\footnote{We estimate the significance by $S / \sqrt{S + B}$,
  where we determine S and B by fitting each peak to a gaussian and counting
  the number of S and B events within twice the fitted width.
  }}. \ However, the signal is very sensitive to the fermion-resonance
  couplings, $\propto g_{ffV}^2$, and thus very suppressed for fermiophobic
  models.
  
  {\bf $s$-channel production to  $W \gamma $: } The second $s$-channel production final state
  we consider is $W^{\pm} \gamma \rightarrow \ell^{\pm} \nu \gamma$. \ Of the
  conventional three vector boson terms the only permutation consistent with
  $U (1)_{em}$ gauge invariance is $g_{\gamma W_{1, 2} W} (\partial_{[\mu,}
  A_{\nu]} (W_{1, 2[\mu}^- W^+_{\nu]}) + h.c.),$ i.e. where the derivative
  acts on the photon field. A nonzero value for only one triboson coupling
  permutation is not possible in traditional, AdS-based Higgsless models.
  However, this final state as been considered recently~{\cite{LH:2007db}} in
  the context of Low-Scale Technicolor (LSTC), exhibiting only one resonance.
  
  The important backgrounds for this process are $W + \gamma$ (irreducible),
  $W + \tmop{jet}$ and $t \bar{t}$ (a jet faking a photon), and $\gamma +
  \tmop{jets}$ (jet fakes \ lepton). We use the rates $\sim 0.1\%$ for a jet to
  fake a photon, and $\sim .02\%$ for a jet to fake an electron
  {\cite{unknown:2006fr}}. We apply the following cuts, 1.) Exactly 1 lepton,
  $p_T > 10 \tmop{GeV}, | \eta | < 2.5$, 2.) Exactly 1 photon, $p_T > 180
  \tmop{GeV}, | \eta | < 2$, 3.) $p_{T, W} > 180 \tmop{GeV}$, 4.) Missing
  Energy $E_{T, \tmop{miss}} > 20 \tmop{GeV}$. See TABLE II for details.
  
  \begin{table}[h]
    $\begin{array}{|c|c|c|c|}
      \hline
      \text{\tmtextrm{Process}} & \sigma_{0, LO}  &
      \epsilon (\%) & N_{EV} ( \mathcal{L} = 10 \tmop{fb}^{- 1})\\
      \hline
      \text{\tmtextrm{HTC1}} & 0.015 \tmop{pb} & 60 & 88\\
      \hline
      \tmop{HTC} 2 & 0.03 \tmop{pb} & 37 & 118\\
      \hline
      W (\ell \nu) + \gamma & 3.84 \tmop{pb} & 0.56 & 215\\
      \hline
      W (\ell \nu) + \tmop{jet} \times \text{\tmtextrm{fake rate}} & 3.06 \tmop{pb} &
      0.28 & 86\\
      \hline
      t \overline{} t (\tmop{leptonic}) & 22.8 \tmop{pb} & 0.005 & 12\\
      \hline
      \gamma + \tmop{jet} \times \tmop{fake} \tmop{rate} & 5.31 \tmop{pb} &
      0.044 & 25\\
      \hline
    \end{array}$ \label{table:wgam}
    \caption{Processes, cross sections and efficiencies for $\tmop{pp}
    \rightarrow \gamma W$ signal and background. }
  \end{table}
  
  Both signals are dramatic, and would be seen within the first few
  $\tmop{fb}^{_{^{- 1}}}$. If the resonances are separated by $\gtrsim 100
  \tmop{GeV}$, as in HTC2, only the lightest resonance is visible because the
  decay modes $W_2 \rightarrow W_1 + Z, Z_1 + W$ are open suppressing the
  branching ratio to $W \gamma$. Conversely, HTC1 exhibits two resonances
  close in mass and both are visible in this channel.
  
  Large $s$-channel signals to $W Z$ and $W \gamma$ are familiar from LSTC {\cite{LH:2007db}}. However, in the PYTHIA implementation of
  LSTC, techni-parity is imposed. Within this approximation only the vector
  resonance couples to $W Z$, and only the axial couples to $W \gamma$.
  This approximation does not hold in the HTC region of interest (viable with
  electroweak constraints).
  
  {\bf $s$-channel production of $Z_{1, 2}$ : }  To discover the neutral
  partners one could use the dilepton or diboson channels: $p p \rightarrow
  Z_{1, 2} \rightarrow \ell^+ \ell^-, W W$. In the NWA, the ratio of dileptons
  to dibosons cross sections is $ \frac{1}{8} \frac{g_1^2}{g_{f f V}^2} \left(  \frac{M_{W_{1, 2}}}{M_W}
  \right)^4 $. For
  $g_{f f V} \simeq g_1$ and a $600 \tmop{GeV}$ resonance
  this ratio is almost 400. However the $W W$ channel may not reveal both
  resonances. In the fully leptonic case $W W \rightarrow \ell \nu \ell'
  \nu'$ we cannot reconstruct the $W W$ invariant mass, while in the semileptonic
  channel $\ell \nu j \nonesep j$ we have to deal with large backgrounds from
  $W + \tmop{jets} \tmop{and} \overline{t} t$. Despite the smaller cross
  section $(\sigma \times \tmop{BR} \sim 1 \tmop{fb})$, the dilepton channel
  may reveal both resonances in the tens of $\tmop{fb}^{- 1}$ range and allow
  for a more detailed study of resonance properties.
  
  {\bf Vector boson fusion: } The second process we investigate is Vector
  Boson Fusion (VBF). VBF will be important at the LHC regardless of the
  fermion-resonance coupling because it provides a window into $W_L W_L$
  scattering. VBF has been studied at parton level for fermiophobic AdS-based
  Higgsless models in Ref. {\cite{hep-ph/0412278,He:2007ge}} where there is
  only one light resonance.
  
  \begin{figure}[t]
    \epsfig{file=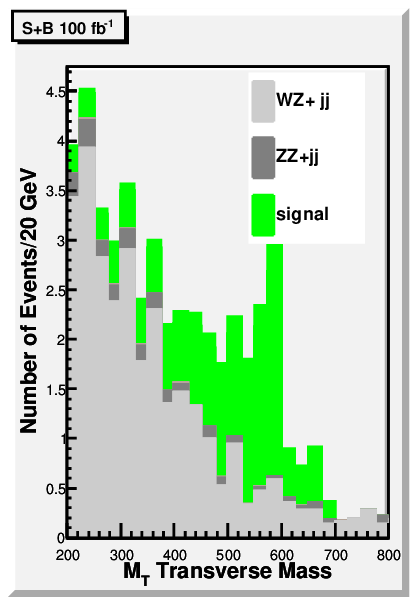,width=1.5in,height=1.25in}\epsfig{file=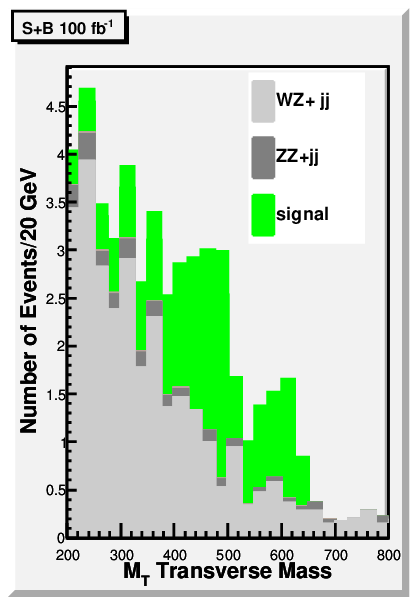,width=1.5in,height=1.25in}\label{VBFfig}
    \caption{VBF channel $p p \rightarrow \tmop{W Z j j}$, ($\mathcal{L}= 100
    \tmop{fb}^{- 1}$) for HTC1 (left) and HTC2 (right). \ }
  \end{figure}
  
  Although new final states can be considered in VBF like $p p \rightarrow
  W_{1, 2} j j \rightarrow W \gamma j j$, we focus on the better studied
  {\cite{Dobado:1990jy,hep-ph/9504426,Dobado:1989ue,Dobado:1990am,Chanowitz:2004gk,hep-ph/0201098}}
  process $p p \rightarrow W_{1, 2} + jj \rightarrow W^{\pm} Z + jj \rightarrow
  3 \ell + \nu + jj$. The backgrounds we consider are $W^{\pm} Z +
  \tmop{jets}, Z Z + \tmop{jets}$. The first
  background is irreducible, while the others come from either missing one
  of the leptons or a jet faking a lepton. \ Our initial cuts are very similar
  to the cuts used in~{\cite{He:2007ge}}: 1.) 3 leptons $p_T > 10 \tmop{GeV},
  | \eta | < 2.5$ 2.) 2 $\tmop{jets}, p_T > 30 \tmop{GeV}, 2 < | \eta | <
  4.5$, 3.) $\Delta \eta_{jj} > 4$, 4.) $|M_{\ell^+ \ell^-} - M_Z | < 7.8
  \tmop{GeV}$. After initial cuts, other backgrounds are negligible. To further
  enhance the significance we also apply a cut $p_{T, Z} > 70 \tmop{GeV}$.
  
  In FIG. \ref{VBFfig} we plot the transverse cluster mass
  {\cite{hep-ph/0508097}} $M_T = ( \sqrt{M^2 (\ell \ell \ell) + p_T^2 (\ell
  \ell \ell)} + |E_{\tmop{missT}} |)^2 - |p_T (\ell \ell \ell) +
  E_{\tmop{missT}} |^2$ . Two edges are visible in HTC2 and one in HTC1. It is
  worth noticing here that PGS tagging efficiency for the VBF forward jets
  ( $\sim 80\%$) is probably too optimistic for LHC at high luminosity.
  
  \begin{table}[h]

    $\begin{array}{|c|ccc|c|}
      \hline
      \text{\tmtextrm{Process}} &  &  & \sigma \times BR \times \epsilon (
      \text{\tmtextrm{final}}) & N_{EV} ( \mathcal{L} = 100 \tmop{fb}^{- 1})\\
      \hline
      \text{\tmtextrm{HTC1}} &  &  & 0.194 \tmop{fb} & 19\\
      \hline
      \tmop{HTC} 2 &  &  & 0.178 \tmop{fb} & 18\\
      \hline
      W Z + \tmop{jets} &  &  & 0.563 \tmop{fb} & 56\\
      \hline
      \tmop{ZZ} + \tmop{jet} &  &  & 0.049 \tmop{fb} & 5\\
           \hline
    \end{array}$

    \caption{Processes and efficiencies for VBF in the $3 \ell+\nu+ 2j$ channel. The following cuts $p_{T,j}>30 $ GeV, $2<|\eta_{j}|<4.5$ and $\Delta R_{jj}>4$ are applied at the parton level. }
  \end{table}

  {\bf Other channels- $W Z Z$, $W \gamma$Z, $W \gamma \gamma$: } Within the HTC
  framework we are also able to study processes with more than two final state
  gauge bosons. These processes have two sources. The first is associated
  production of a $Z$ (or $\gamma$) with a resonance which subsequently decays
  into $W^{\pm} Z (\gamma)$: $p p \rightarrow Z W_{1, 2} \rightarrow
  W Z Z, \tmop{etc} .$ The second source is the direct production of a
  heavy resonance which has sufficient phase space to decay into a lighter
  resonance plus a SM gauge boson: $p p \rightarrow W_2^{\pm} \rightarrow
  W_1^{\pm}$Z. Associated production is familiar from fermiophobic Higgsless
  models {\cite{hep-ph/0412278,He:2007ge}}. Parton level studies have been
  performed in the $W Z Z \rightarrow 4 \ell j j$ channel, and although
  promising at parton level, the signal deteriorates once showering and
  detector effects are included. We estimate that 200-300 $\tmop{fb}^{- 1}$ of
  luminosity is required for discovery and leave a more thorough study for
  future work.
  
  {\bf Conclusions: } To attack the parameter space of strong EWSB and perform LHC phenomenology, we use an economical parametrization of resonance interactions: Holographic Technicolor (HTC). The role of the 5D modelling here is to reduce the number of parameters to four, down from the ${\cal O}(100)$ couplings possible in an effective Lagrangian of (two triplets of spin-1) resonances coupled to electroweak fields. HTC also allows us to effectively describe situations departing from rescaled QCD, as required by electroweak constraints.

These constraints still allow for light (500 or 600 GeV) resonances close by in mass (100 or 150 GeV separation) and sizeable direct couplings to SM fermion. We performed a collider study of two sample points  (HTC1 and HTC2) in the parameter space of HTC. Both sample points exhibit early discovery of two light and nearby resonances via $s$-channel production to $W Z$ and $W \gamma$. The resonances also appear in VBF, though this requires higher luminosity.

In the future, we intend to provide a
   package based on MADGRAPH-BRIDGE to
  allow users to further navigate the parameter space. The framework
  presented here can be extended to add new particles,
  e.g. techni-pions, techni-omegas and composite Higgs.

   {\bf Acknowledgements: } We thank T. Appelquist, G. Azuelos, G. Brooijmans,
  K. Lane, R. S. Chivukula, N. Christensen, M. Perelstein and W. Skiba for helpful comments. The work of JH and AM is supported by DOE grant DE-FG02-92ER-40704
  and VS is supported by DE-FG02-91ER40676.
  
\bibliography{benchmark}

\end{document}